# Data-generating process and time-series asset pricing

Shuxin GUO
*School of Economics and Management*
*Southwest Jiaotong University*
*Chengdu, Sichuan, P. R. China*
shuxinguo@home.swjtu.edu.cn
ORCID: 0000-0001-8188-6517

Qiang LIU*
*Institute of Chinese Financial Studies*
*Southwestern University of Finance and Economics*
*Chengdu, Sichuan, P. R. China.*
qiangliu@swufe.edu.cn
ORCID: 0000-0001-8466-3108
* corresponding author.

**Acknowledgements**

This work was supported by the National Natural Science Foundation of China under Grant numbers 71701171, 72073109, and 72071162, and by the Liberal Arts and Social Sciences Foundation of the Chinese Ministry of Education (21XJC790003).

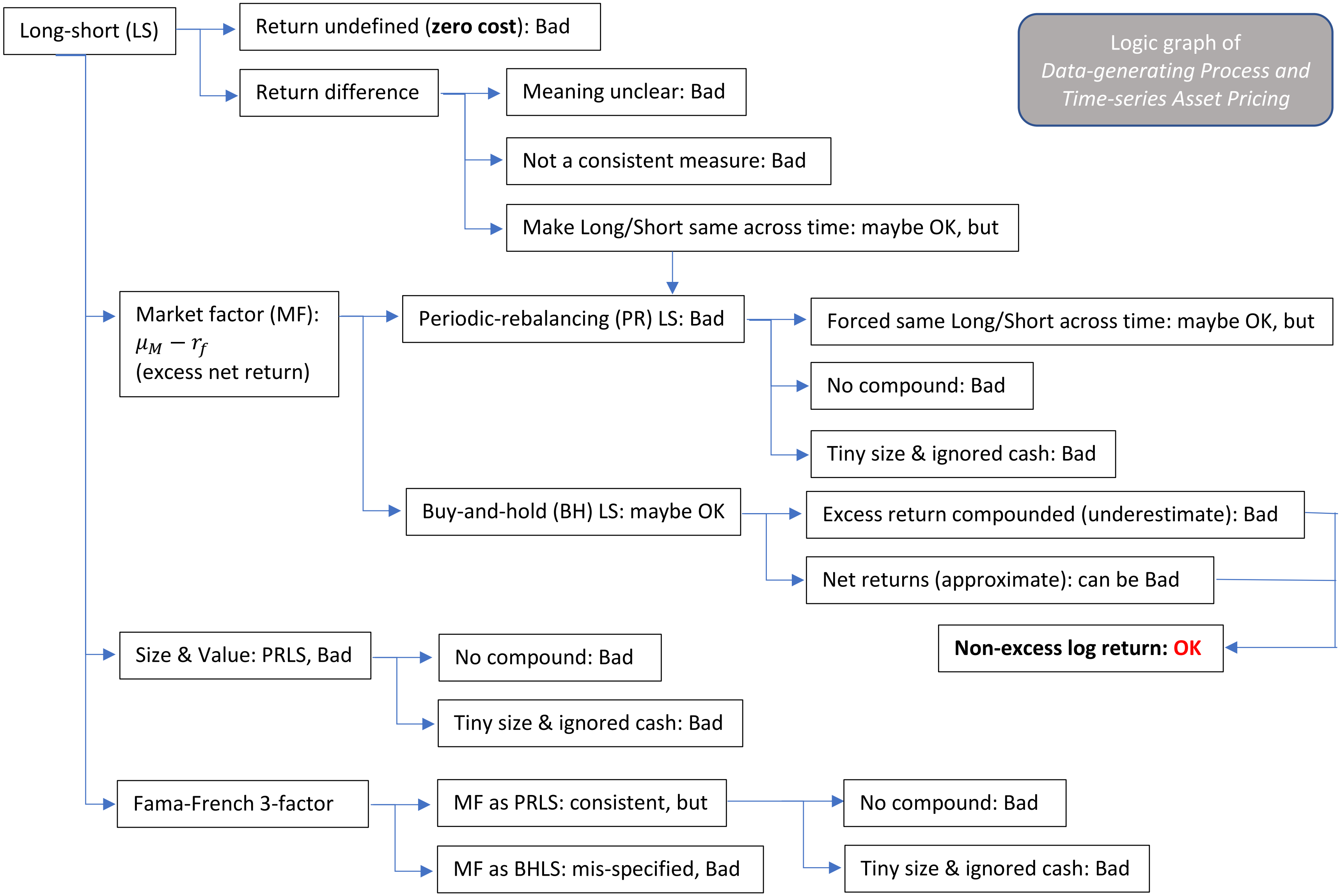

Long-short (LS)
Return undefined (zero cost): Bad
Return difference
Meaning unclear: Bad
Not a consistent measure: Bad
Make Long/Short same across time: maybe OK, but
Logic graph of
Data-generating Process and
Time-series Asset Pricing
Market factor (MF):
$\mu_M - r_f$
(excess net return)
Periodic-rebalancing (PR) LS: Bad
Forced same Long/Short across time: maybe OK, but
No compound: Bad
Tiny size & ignored cash: Bad
Buy-and-hold (BH) LS: maybe OK
Excess return compounded (underestimate): Bad
Net returns (approximate): can be Bad
Non-excess log return: OK
Size & Value: PRLS, Bad
No compound: Bad
Tiny size & ignored cash: Bad
Fama-French 3-factor
MF as PRLS: consistent, but
MF as BHLS: mis-specified, Bad
No compound: Bad
Tiny size & ignored cash: Bad

**Data-generating Process and Time-series Asset Pricing**

**Abstract:** We study the data-generating processes for factors expressed in return differences, which the literature on time-series asset pricing seems to have overlooked. For the factors' data-generating processes or long-short zero-cost portfolios, a meaningful definition of returns is impossible; further, the compounded market factor (MF) significantly underestimates the return difference between the market and the risk-free rate compounded separately. Surprisingly, if MF were treated coercively as periodic-rebalancing long-short (i.e., the same as size and value), Fama-French three-factor (FF3) would be economically unattractive for lacking compounding and irrelevant for suffering from the small "size of an effect." Otherwise, FF3 might be misspecified if MF were buy-and-hold long-short. Finally, we show that OLS with net returns for single-index models leads to inflated alphas, exaggerated *t*-values, and overestimated Sharpe ratios (SR); worse, net returns may lead to pathological alphas and SRs. We propose defining factors (and SRs) with non-difference compound returns.

## 1. Introduction

The mainstream empirical asset pricing over the last half a century try to explain equity excess returns via common factors (e.g., Fama and French (1993) and (2015), Hou et al. (2015), and Liu et al. (2019)) or document new factors or anomalies (see excellent reviews in Hou et al. (2015) and Harvey et al. (2016)). However, the vast majority of such research seems to neglect one critical issue underlying those factors: their data-generating processes. This issue is important, because a clear and better understanding of the data-generating process of a factor can help us avoid model misspecifications and further provide a firm foundation for explaining excess returns. Therefore, we primarily study the data-generating processes for several common factors expressed in return differences in this paper.

The return on assets is undoubtedly the most widely-used concept in financial economics. Investment performance is foremostly gauged by returns, and the Nobel prize-winning capital asset price model (CAPM) is also formulated on returns (Sharpe, 1964). It is well-known that for an investment period, the percentage return can be trivially computed by subtracting one from the ratio of the final value (or price) to the initial value (or price) of an investment. Such a percentage return is called net return (Cochrane, 2005). Net returns are predominantly used in the empirical time-series asset pricing literature so that the popular CRSP includes pre-computed monthly net returns of US stocks in the database. Also, Kenneth French provides monthly net-return differences online for the Fama-French three-factor (FF3) model (Fama and French, 1993), among others.[1]

[1] mba.tuck.dartmouth.edu/pages/faculty/ken.french/Data_Library/f-f_factors.html.

However, the implicit or more fundamental variable behind asset returns is the value of (investment) assets, portfolios, or strategies. As a result of the realization of the data-generating process, the time series of portfolio values is what we can observe directly. Therefore, empirical asset pricing studies should begin with the analysis of values or the data-generating process instead of the return process.

Generally speaking, there are two common categories of value processes. The first is the buy-and-hold process, consisting of the value of a portfolio or the price of a stock across time, which can be approximated by geometric Brownian motions (GBM) as the data-generating process, the simplest continuous-time stochastic process widely used in finance (Merton, 1992; Hull, 2015). The second is a hypothetical periodic-rebalancing process, which starts each period with the same amount of capital. Unfortunately, periodic rebalancing cannot be easily associated with a known stochastic process.[2]

In our comprehensive analyses, we show that the average behavior of a buy-and-hold portfolio is measured correctly only by the average of compound or log returns rather than net returns. Further, buy-and-hold and log returns take the compounding effect into consideration inherently, while net returns do not (see Section 2). Therefore, buy-and-hold is economically meaningful and worthy of study. In contrast, periodic rebalancing is much less worthwhile economically because the gain or loss of it in each period is realized (and then ignored to some extent), and no compounding effect is accounted for.

[2] See Section 2.1 for the definitions of buy-and-hold and periodic-rebalancing.

Fortuitously, our attention to the compounding effect is corroborated by Nobel laureate Myron Scholes. Reportedly, Scholes said that "compound returns were very different to the average returns, or Sharpe ratios, that investors measure" and urged "[i]nvestors need to be patient and harvest the power of compound returns."[3] It is clear that by "average returns," as well as "Sharpe ratios," Scholes is referring to mean net returns that are prevalent in the asset pricing literature. We agree that the compounding effect is of utmost importance in investment and should not be ignored in empirical time-series asset pricing.

Based on buy-and-hold and periodic-rebalancing, we next analyze the data-generating process for long-short portfolios. The market factor is commonly defined by the excess return of the market portfolio (see Section 4.1), and thus either buy-and-hold long-short[4] (BHLS) or periodic-rebalancing long-short (PRLS) (see Section 2.1). The size and value factors in FF3 are by definition PRLS portfolios (see Section 4.2). We argue that the long-short (zero-cost) portfolio should be avoided by asset pricing studies, because a meaningful definition of returns for its value process, which has an initial value of zero, is impossible (see Section 3.1). Further, the market factor or excess market return, which is commonly utilized in regressions, significantly underestimates the compounding effect of the difference between the market portfolio and the risk-free interest rate compounded separately within a BHLS portfolio (see Section 3.4).

[3] Sarah Rundell, 2018, Scholes' strategies for compound returns. https://www.top1000funds.com/2018/10/scholes-strategies-for-compound-returns/, October 3, 2018.

[4] For example, Moreira and Muir (2017) form their managed portfolio on "the buy-and-hold portfolio excess return" in their Equation (1).

Because the market portfolio is naturally buy-and-hold and the OLS regression deals with averages of returns, we argue that only log returns yield the correct average for the market portfolio. Therefore, CAPM and other single index models are correctly formulated only with non-excess log returns. Counterintuitively, the market factor might coercively be viewed as a PRLS portfolio. However, as was pointed out earlier, periodic rebalancing does not consider the compounding effect and, thus, is economically less relevant. Worse, periodic rebalancing suffers from what we call "the tiny size paradox" (see Section 5.1), because the equal amount of capital after each rebalancing has to be fixed to a tiny value, which can only be a small percentage of the initial market capitalization (for being tradable and not moving the market simultaneously) and might have been determined up to many decades ago.[5] The tiny size is no doubt an example of the small "size of an effect," which is deemed economic irrelevancy (Nuzzo, 2014).[6]

Unlike the marker factor, the well-known size and value (i.e., book-to-market) factors in FF3 are defined as zero-cost PRLS portfolios. They cannot be modeled as continuous-time stochastic processes for sure in the first place. Consequently, the FF3 model can be considered in two different fashions. If the market factor were buy-and-hold, the three Fama-French factors would be inconsistent. Arguably, equity excess returns in asset pricing, which are buy-and-hold in nature (with compounding) with the corresponding stock value usually growing over time, cannot be explained by PRLS factors that lack compounding and remain fixed to the initial value (i.e., amount of capital). Therefore, the FF3 model might be misspecified. On the other hand, if the market factor were coercively regarded as periodic-rebalancing, the FF3 model would seem to be specified consistently. But in this case, the model would become economically insignificant and irrelevant for two reasons once again. First, periodic rebalancing does not take compounding effects into account and, therefore, cannot "harvest the power of compound returns" as urged by Scholes. Second, the three Fama-French factors together will suffer from "the tiny size paradox" even more severely than the market factor alone. The total capital of size and value, which accounts for only 7% of the market capitalization in 1926, to begin with, becomes minuscule in 2020 for being less than 0.005% of the US market (see Section 5.1). Therefore, we suggest that the asset pricing factors would be better represented by non-excess or non-difference log returns. Importantly, our analyses also apply to other FF3-type models, such as the Fama-French five-factor (Fama and French, 2015), the $q$-factor model (Hou et al., 2015), and the Chinese three- and four-factors (Liu et al., 2019), among others.

[5] For example, the market portfolio in Moreira and Muir (2017) goes back to 1926.
[6] Tables 4 and 5 later show that the US market capitalization in 1926 is less than 0.07% of that in 2020.

One might argue that (monthly) net returns can approximate log returns. For example, however, the approximation turns out very bad for the S&P 500 index. The difference between the monthly means of net and log returns from January 1, 1960 to January 1, 2020 is only 0.0009 and is really tiny. Unfortunately, if compounded, the mean of net returns overestimates the index's terminal value by 89%, an egregious error that cannot be neglected. Through numerical analyses, we further show that using net returns in OLS will result in inflated $t$-values, exaggerated alphas, and overestimated Sharpe ratios. In certain extreme circumstances, net returns lead to alphas with wrong signs and Sharpe ratios with incorrect ordering and, thus, are not guaranteed to be correct. Based on these findings, we suggest that both the market factor and Sharpe ratio should be defined with (buy-and-hold) log returns instead of excess net returns.

The remaining body of the paper is organized as follows. In Section 2, we define the buy-and-hold strategy and the hypothetical periodic-rebalancing strategy. Further, we analyze the geometric and arithmetic means of returns and show that the geometric mean of returns is the correct average for buy-and-hold time series. The long-short portfolios, ubiquitous in asset pricing, are discussed in Section 3. Section 4 analyses the value processes of the single-index model and the FF3 model. We present additional numerical analyses of the US market and simulated buy-and-hold strategies in Section 5. Section 6 concludes.

## 2. Portfolio, Return and Mean Return

In this section, we mainly focus on the data-generating processes of several possible portfolios. In investing and finance, data-generating processes are observed for or associated with the value processes of tradable portfolios or strategies. In asset pricing, however, the averages of returns are employed instead (see, for example, Fama and French (1993)). To make the exposition self-contained, we first reintroduce (or define) three widely-used portfolios in the literature and then discuss their corresponding data-generating processes. Because the averages of returns other than (data-generating) value processes are employed in asset pricing, we also discuss three returns and their means and, finally, the relationships between value processes and their corresponding means of returns.

### 2.1. Three portfolios

Essentially, time series empirical asset pricing studies deal with portfolios or strategies. In this section, we first reintroduce three prevalent portfolios in the literature, namely buy-and-hold (BH), periodic-rebalancing (PR), and zero-cost (ZC), for easy exposition later and completeness. We then focus on analyzing the underlying data-generating processes for these portfolios.

**Buy-and-hold** If we buy shares of a stock and hold them for a number of years, we are essentially utilizing the buy-and-hold strategy. The value of a portfolio, or more simply the price of a stock, across time, is then a buy-and-hold price process. In finance, buy-and-hold price processes are modeled by continuous-time stochastic processes as the corresponding data-generating processes, with geometric Brownian motions, Ito processes, and stochastic volatility models commonly employed to model the dynamics for stocks and other underlying variables (Merton, 1992; Hull, 2015). Moreover, being composed of all public stocks, the (value-weighted) market (portfolio) is also a buy-and-hold portfolio by definition (for example, see Moreira and Muir (2017)).

**Periodic-rebalancing** For comparison, we imagine a hypothetical strategy in which the investing process starts each period with the same amount of capital. It will be called the periodic-rebalancing strategy in this paper. Unfortunately, it is difficult to associate the periodic-rebalancing value series with a known stochastic process.

Further, the meanings of mean and variance for periodic-rebalancing may behave strangely, as we will show later in Section 3.

The difference between buy-and-hold and periodic-rebalancing strategies is as follows. With a buy-and-hold strategy, the gain (loss) in each period is not realized until the portfolio is liquidated at the end of an investment. In popular media, the unrealized cash is called a paper gain (loss). In the periodic-rebalancing strategy, investors take out or realize the gain (loss) in each period for sure so that the capital is readjusted to be the same again at the beginning of the next period. Consequently, the periodic-rebalancing strategy is associated with an extra cash process, which by itself has the time value of money and can earn risk-free rates at least. But unfortunately, this important issue is ignored by most studies in the literature. With extra cash flows, periodic rebalancing becomes more complex than the buy-and-hold strategy to analyze. Further, one might be tempted to believe that the cash flows are insignificant with respect to the final total return. Unfortunately, this is incorrect in fact. Section 2.4 later shows that the cash flows compounded on risk-free rates easily dominate the simple sum of realized returns in each period.

Notably, the difference between buy-and-hold and periodic-rebalancing has critical ramifications financially. The buy-and-hold strategy accumulates or compounds the gain (loss) in each period, while the periodic-rebalancing strategy lacks the effect of compounding. In long-term investing, however, the compounding effect is usually the driver of superb returns. Therefore, the strategies or factors related to buy-and-hold are arguably more important and practically more relevant to study.

**Long-short** The long-short zero-cost portfolio, commonly employed in empirical asset pricing, is set up initially by selling short a certain amount of securities and using the whole proceeds to buy other securities.[7] Later, we can choose to make it zero-cost again or not. To make it zero-cost at the start of each period and maintain the same long/short amount across time,[8] we will have to adjust the short position or long position or both, which is referred to as portfolio-rebalancing, as was discussed in the immediate previous section. Also, the adjustments lead to extra cash flows, positive or negative, just as in the case of periodic-rebalancing. Consequently, such a portfolio suffers the same drawbacks as a periodic-rebalancing portfolio. For easy reference, we call it periodic-rebalancing long-short (PRLS). From the trading point of view, PRLS is equivalent to the difference between two periodic-rebalancing portfolios. Note that the size and value factors in FF3 are actually periodic-rebalancing long-short (zero-cost) portfolios (see Section 4 later).

If the initial zero-cost portfolio is not rebalanced later, we end up with two buy-and-hold portfolios: one short and one long. Let's call it buy-and-hold long-short (BHLS). In fact, the market factor in FF3 can be either BHLS or PRLS.

By definition, the total value of PRLS begins each period with zero. On the other hand, the total values of BHLS can be zero (at the beginning of the course, but later as well), positive, or negative. Consequently, a reasonable return series (or data-generating process) cannot be defined for long-short portfolios, given that zero or negative values are involved. This makes long-short portfolios hard to analyze, as we will discuss later in Section 3.

[7] In real-world trading, zero-cost is only possible for borrowing cash (to long risky securities), because a margin of at least 50% is required to sell short and the whole short-sale proceeds have to be kept in the margin account. In addition, short-sellers may face margin calls and have to put in extra capitals. This implies that long-short zero-cost, as is discussed in asset pricing, is not tradable in practice. For simplicity, we do not discuss this implication here further, though.

[8] In periodic-rebalancing, the average of time-series returns is meaningless, if the initial capitals are not the same across time. For example, the average return of 10% is clearly wrong for 30% year-1 return with capital 10 and -10% year-2 return with capital 1000, because the two-year investment actually incurs a total loss of 97.

### 2.2. Three returns

For easy exposition later, we reintroduce gross return, net return, and log return here. Superficially, they are well-known and seem to be trivial concepts that are not worth any further discussions. As we will see, however, the averages or means of those returns are dramatically different. Therefore, let's be familiar with them anyway.

Assume that we have a stock price process, $\{S_t, t = 0,1,,\cdots,n\}$. If $S_t$ is regarded as the price of a stock or the value of a portfolio at time *t*, we would then have a buy-and-hold portfolio. As a result, we call the set $\{S_t\}$ a buy-and-hold price process.

**Gross return** The ratio of two prices is called the gross return as in the following:

$$R_t = \frac{S_t}{S_{t-1}}$$

**Net return** The gross return minus one is the net return:

$$\mu_t = R_t - 1$$

**Log return** Finally, the continuously-compounded or log return is as follows (Cochrane, 2005):

$$y_t = lnR_t$$

### 2.3. Two means of returns

Empirical asset pricing mainly deals with the mean and standard deviation of returns

(see Formula (6) later). Mathematically, a mean can be either arithmetic or geometric, depending on the definitions of returns.

**Geometric mean of returns** The geometric average of gross returns over an *n*-period for a buy-and-hold strategy can be called the geometric mean of returns:

$$\bar{R}_g = \left(\prod_{t=1}^{n} R_t\right)^{\frac{1}{n}} = \left(\frac{S_n}{S_0}\right)^{\frac{1}{n}}$$

With a root of $\frac{1}{n}$, $\bar{R}_g$ is difficult to work with. Fortunately, though, the $\bar{R}_g$ formula above is equivalent to the arithmetic average of log returns as in the following:

$$\bar{y} = \frac{1}{n}\sum_{t=1}^{n} y_t = \frac{1}{n}\ln\left(\frac{S_n}{S_0}\right) \quad (1)$$

where $\bar{y} = ln\bar{R}_g$ and $y_t = lnR_t$. It is worth noting that $\bar{y}$ depends only on the initial price and terminal price and thus is path-independent. Similarly, $\bar{R}_g$ is also path-independent.

The nature of path independence of the geometric mean of returns is vitally important. Due to path independence, we can arrive at $S_n$ from $S_0$ via either $\bar{R}_g$ or $\bar{y}$:

$$S_n = S_0\bar{R}_g^n = S_0e^{n\bar{y}}$$

Therefore, $\bar{R}_g$ or $\bar{y}$ correctly describe the compounded "average return" over each period for the buy-and-hold price process.

It is important to note that the arithmetic average of log returns, which is easier to deal with than the geometric average of gross returns, nevertheless maintains the nature of the geometric mean or effect of compounding. Therefore, we can simply refer to $\bar{y} = ln\bar{R}_g$ as the geometric mean of returns.

The path independence of the geometric mean of returns results from the nature of the buy-and-hold process, in which the end value of a period is the beginning value of the immediate next period. When the value process is not continuous, or the end value of a period is different from the beginning value of the immediate next period, the geometric average of gross returns is not well defined. Therefore, the geometric average does not measure the correct "average" of a discontinuous value process. This surprising conclusion implies that many of the reported compound average returns for hedge funds or mutual funds are incorrect because funds frequently have cash inflows or outflows (Liu and Guo, 2022).

**Arithmetic mean of returns** For convenience, the arithmetic average of net returns over the *n*-period is referred to as the arithmetic mean of returns:

$$\bar{\mu} = \frac{1}{n}\sum_{t=1}^{n} \mu_t \tag{2}$$

which is equivalent to the arithmetic average of gross returns as follows:

$$\bar{R}_a = \frac{1}{n}\sum_{t=1}^{n} R_t$$

where $\bar{R}_a = 1 + \bar{\mu}$ and $R_t = 1 + \mu_t$. Unlike the path-independent $\bar{R}_g$, $\bar{R}_a$ does depend on all the stock prices along a stock path or in a time series and therefore is path-dependent. Similarly, $\bar{\mu}$ is also path-dependent.

The path dependence of the arithmetic mean of returns makes it much less useful. It is impossible to reach $S_n$ from $S_0$ via $\bar{\mu}$ (or $\bar{R}_a$) without the complete price details in between. The reason is as follows. There are only two ways, either with or without compounding, to go from $S_0$ to $S_n$. With compounding, we have:

$$S_0(1+\bar{\mu})^n = S_0\bar{R}_a^n > S_0\bar{R}_g^n = S_n$$

due to the well-known arithmetic mean-geometric mean inequality, or $\bar{R}_a > \bar{R}_g$. In other words, the arithmetic mean will exaggerate the true return or growth of a buy-and-hold process if employed under compounding.

Without compounding, the gain (loss) could be $\bar{\mu}S_0$ in each period, provided we start each period with the same amount of $S_0$. Critically, let's do not forget that the cash can earn (or has to be financed with) risk-free interest rates with compounding for simplicity.[9] Thus, the terminal value:

$$S_0 + \bar{\mu}S_0 \sum_{t=1}^{n-1} \prod_{s=t}^{n-1} (1+r_s) + \bar{\mu}S_0 \tag{3}$$

is not equal to $S_n$, where $r_t$ is the one-period risk-free interest rate at the beginning of period *t*. Note that the cash deposit or loan in each period is always compounded in practice. Therefore, the arithmetic mean of returns is not a correct measure of the "average return" over each period for the buy-and-hold price process, with or without compounding.

More surprisingly, the arithmetic mean of returns is not a correct measure of the "average return" over each period for the periodic-rebalancing price process, either. The terminal value of the periodic-rebalancing strategy is actually:

$$S_0 + S_0 \sum_{t=1}^{n-1} \mu_t \prod_{s=t}^{n-1} (1+r_s) + \mu_n S_0 \tag{4}$$

which is more than $S_0(1+n\bar{\mu})$. Note again that the cash (gain or loss) in each period can earn (or has to be financed with) risk-free interest rates. Expression (4) is obviously not equal to Expression (3); worse, equality does not hold even if the risk-free rate is constant.

[9] This can be assumed for institutional investors, even though the real situations are more complicated.

This peculiar dependence of the arithmetic mean of returns on the initial amount of capital arises from the asymmetry between negative and positive net returns. This can be seen more easily from numerical examples with extreme values. For a price process of {100, 50, 100}, $\mu_1 = -50\%$ and $\mu_2 = 100\%$. Clearly, a higher net gain is needed to cover a previous net loss; the simple (arithmetic) average of the two net returns, which is 25%, is no doubt absurd.

At worst, a stock can only lose -100% under net return when its price drops to zero. In contrast, the potential upside gain is essentially unlimited and infinite. This asymmetry implies that negative and positive net returns should not be regarded as equivalent in a time series, and we cannot obtain the correct average of a net return series if the initial amounts in each period are different. Therefore, the arithmetic mean of returns is economically meaningless in general. But fortunately, the log returns are symmetric, and the geometric mean of returns measures the "average return" of a buy-and-hold strategy correctly.[10]

Importantly, this critical insight carries over and applies to zero-cost portfolios. For example, assume that we start with \$100 long and \$100 short for the first period. With a net (long-short) loss of -50%, the zero-cost investment will lose \$50. If we then begin the second period with \$50 long and \$50 short, we again need a net (long-short) gain of 100% to earn \$50 and cover the loss in the first period. Similarly, the arithmetic mean of returns is economically meaningless for zero-cost portfolios in general.

[10] With log returns, $y_1 = -ln2$ and $y_2 = ln2$. The mean of $y_1$ and $y_2$ is zero and thus correct.

This is a vitally important point worthy of emphasizing. Given a buy-and-hold price process, we can easily compute the corresponding net-return series. Further, we can calculate the mean and other statistics of $\mu_t$ mathematically. Those statistics are path-dependent, though. They do not represent the true average behaviors of unbiased random paths and, in general, are less meaningful economically, however. Therefore, the usefulness of net returns can be extremely limited for time series empirical studies.

**2.4. S&P 500 index**

This section applies numerical analyses to make the above concepts more readily understandable. We use the S&P 500 index as a proxy for the US market. The monthly adjusted closes from January 1, 1960 to January 1, 2020 are downloaded from Yahoo!Finance. For easy exposition, we present the analyses of the S&P 500 index in Table 1.

Table 1 Here

The index started at 55.61 on January 1, 1960, and finished at 3225.52 on January 1, 2020, covering 720 months or sixty years. For studying buy-and-hold strategies, a sixty-year investment is arguably quite long, even though recent literature on empirical asset pricing routinely covers about ninety years.

The comparison between log returns and net returns is striking. Net returns overestimate the mean but underestimate the standard deviation. The differences appear quite small, but the Sharpe ratio under net returns is exaggerated by nearly 30%. This implies that when employed in regressions, net returns will exaggerate the alpha and its *t*-value and lead to more false discoveries. It is worth noting that the difference between the arithmetic and geometric monthly means is only 0.0009 and really tiny. One may be tempted to argue that the arithmetic mean (or the net return) approximates the geometric mean (or the log return) well. Unfortunately, the error of approximation by the arithmetic mean, which inflates the geometric mean by 16%, leads to egregious errors when compounded, as we will see in the next paragraph.

As expected, we arrive at the correct terminal value via the geometric mean from the initial value. Thus, the geometric mean of returns does measure the average return of the buy-and-hold portfolio. In contrast, the arithmetic mean, if compounded, overestimates the terminal value by 89% and is unsatisfactory. Consequently, it is a wrong measure for the average return of the buy-and-hold portfolio.

The 720 realized returns on the initial value of 55.61 sum up to only 261.99, while the corresponding terminal value of 317.60 is less than 10% of the actual terminal value of 3225.52. The realized cash in each period adds up to 674.17 (951.20) if compounded to the end of investment with a constant risk-free rate of 3% per annum (the actual one-month treasury bill returns[11]), about 2.6 (3.6) times the simple sum (i.e., 261.99). Clearly, the compounding effect is the main driver of the risk-free investment returns of the realized cash in a buy-and-hold fashion. Note unfortunately that the cash realized on the arithmetic mean compounded with a constant 3% rate (the actual one-month treasury bill returns), which sum up to 733.01 (1170.24), over-estimate the final investment (i.e., 674.17 (951.20)).

[11] The returns are from the data library of K. French.

Overall, three observations can be made. First, the arithmetic mean is not a correct measure even for the average return of the periodic-rebalancing strategy. Second, the primary gain of the periodic-rebalancing strategy comes from the cash compounded by the risk-free rates, which have been ignored in the literature. This implies that it is economically irrelevant to study periodic-rebalancing portfolios while neglecting the gain or loss in each period in empirical asset pricing. Third, the buy-and-hold strategy significantly outperforms the periodic-rebalancing strategy by taking the compounding effect into consideration.

## 3. PRLS and BHLS

The size and value factors in FF3 are periodic-rebalancing long-short (PRLS) portfolios, as will be discussed in Section 4. The buy-and-hold long-short (BHLS) portfolio mentioned in Section 2.1 may appear exotic but is arguably employed in empirical asset pricing. The market factor in both CAPM and FF3 is defined by the difference between the market net return and the risk-free interest rate (i.e., the excess market return. see Equation (5) later). Implicitly but essentially, the definition implies either a PRLS or a BHLS portfolio with the market as the long asset and the borrowed cash (or a risk-free bond) as the short asset. Due to the ubiquity of long-short portfolios, it is crucial to understand their nature, which we analyze in this section.

### 3.1. Long-short portfolio return

Assume that we borrow 100 and use it to buy a stock; at the end of one year, the stock grows to 110, and we make seven dollars after we sell the stock and repay the loan at 103. For easy exposition, we denote the long-short price-pair process as $\begin{Bmatrix} 100 \\ 100 \end{Bmatrix}, \begin{Bmatrix} 110 \\ 103 \end{Bmatrix}$, and the portfolio value process as {0, 7}. With an initial value of zero (i.e., zero-cost), a rational return cannot be defined for the value process, unfortunately. With net returns $\mu_L = 10\%$ and $\mu_S = 3\%$, asset pricing treats the net-return difference $\mu_L - \mu_S = 7\%$ as the portfolio return.[12] Then, a critical question can be asked: What, the initial value of zero or the short amount of 100, is the 7% return on? In other words, the economic meaning of the long-short portfolio return employed in asset pricing is ambiguous at best.

Worse, return differences can lead to contradictions if initial short values are different. Let's see the following inset:

| Price-pair process | Value process | Return difference |
|---|---|---|
| $\begin{Bmatrix} 1000 \\ 1000 \end{Bmatrix}, \begin{Bmatrix} 1100 \\ 1030 \end{Bmatrix}$ | $\{0,70\}$ | 7% |
| $\begin{Bmatrix} 100 \\ 100 \end{Bmatrix}, \begin{Bmatrix} 115 \\ 103 \end{Bmatrix}$ | $\{0,12\}$ | 12% |

Obviously, process {0, 70} with 7% contradicts both process {0, 12} with 12% as well as process {0, 7} with 7%. So, let's summarize the results in a proposition.

**Proposition 1** For a long-short portfolio with the price-pair process $\begin{Bmatrix} S_0 \\ S_0 \end{Bmatrix}, \begin{Bmatrix} L_1 \\ S_1 \end{Bmatrix}$ and the value process $\{0, L_1 - S_1\}$, the net-return difference is defined as $\mu_{LS} = (L_1 - S_1)/S_0$.

(a) A meaningful return cannot be defined for long-short portfolios.

(b) The net-return difference, commonly regarded as the portfolio return, is not a

[12] A return difference implies that the long and short legs have the same initial capital to begin with. If the initial long and short capitals differ, it is unclear how a return for the portfolio can be defined and the simple return difference will be meaningless.

consistent measure for the portfolio value process.

Therefore, the long-short portfolio, without a meaningful definition of returns, is of limited value and should be avoided by asset pricing studies. However, one might argue that we may still treat the net-return difference as the portfolio return as long as we keep the initial price-pair constant across time (a.k.a., PRLS). In this way, the net-return difference is at least a consistent measure, even though its exact meaning may be unknown. So, let's analyze PRLS next.

**3.2. PRLS mean**

Given the net-return difference as the portfolio return, we now analyze periodic-rebalancing long-short (PRLS) portfolios. Because PRLS starts each period with $\{\frac{S_0}{S_0}\}$, we only show the initial pair once in the following price-pair process:

$$\left\{\frac{S_0}{S_0} :: \frac{L_1}{S_1}, \frac{L_2}{S_2}, \dots, \frac{L_n}{S_n}\right\}$$

With net returns, it is easy to show that the mean of net-return differences is

$$\bar{\mu}_{LS} = \frac{1}{S_0}(\bar{L} - \bar{S})$$

where $\bar{X} = \frac{1}{n}\sum_{t=1}^{n} X_t, X = L, S$. Further, the variance of net-return differences is

$$\sigma_{\mu_{LS}}^2 = \frac{1}{S_0^2}(\sigma_L^2 + \sigma_S^2 - 2cov[L, S])$$

where $\sigma_{\mu_{LS}}^2 = E[\mu_{LS} - \bar{\mu}_{LS}]^2$, $\sigma_X^2 = E[X - \bar{X}]^2, X = L, S$, and $cov[L, S] = E[(L - \bar{L})(S - \bar{S})]$.

The mean of net-return differences or the PRLS mean is essentially the difference of price means. The variance of net-return differences, depending on the covariance of two end-period price series, differs from the simple sum of variances of two ending-period price series. Viewed via trading, PRLS is nothing but the difference between two periodic-rebalancing portfolios, and the total variance of PRLS would be the sum of the variances of the two portfolios (i.e., the sum of variances of two ending-period price series). In other words, the net-return difference series is not the same as the difference between the two net-return series. Finally, PRLS net-return differences might be problematic when used in regression because level price series are known to show spurious correlations. The above discussion implies that log-return differences, which do not accurately describe the PRLS value process due to Proposition 1a either, may be better than net-return differences in approximating PRLS.

### 3.3. BHLS mean

The buy-and-hold long-short (BHLS) portfolio, composed of both a long buy-and-hold asset and a short buy-and-hold asset, initially has a setup value (investment) of zero but can later have either positive or negative values. With a starting value of zero, the whole BHLS portfolio also forbids a meaningful definition of returns. Further, a return cannot be reasonably defined if an investment's initial or final value is negative. Therefore, as in the case of PRLS, we have to treat the long and short value series separately and describe the return series of the portfolio approximately.

As discussed previously, the average behavior of a buy-and-hold portfolio can only be measured correctly by log returns. It turns out that the average behavior of a BHLS portfolio is approximated better by log returns, too. We use a two-period BHLS portfolio to make the point clear.

Table 2 shows three hypothetical scenarios for BHLS portfolios. In Panel A, BHLS ends up with no gain or loss. With log return, the difference between Long (Short) buy-and-hold is 62.9% (-62.9%), leading to the correct mean of zero. In contrast, the corresponding mean from net returns is 5.8%, which is undoubtedly incorrect. Panel B shows a final positive gain of 45 for BHLS. Even though it is impossible to define returns for the BHLS value series of {0, 85, 45}, the mean return has to be positive somehow. Therefore, 24.8% from log returns has the correct sign, while -246.1% from net returns does not. Finally, BHLS in Panel C shows a final loss of -10. Similarly, the mean from log returns is -6.7% with the correct sign, but that from net returns is 235.0% and once again with the wrong sign.

Table 2 Here

Granted, the above scenarios in Table 2 are somewhat contrived. Still, we can safely conclude from those numerical results that log returns better approximate the average behavior of BHLS.

**3.4. Excess return**

If we treat the market factor as a BHLS portfolio, the analysis of the BHLS mean applies similarly. To make the discussion a little more realistic, we present simulated results in the following.

Table 3 shows two simulated paths for one risky asset. The value of the risky asset follows GBM with an expected annual (arithmetic) mean of 10% and annual volatility of 30%. The risk-free asset grows continuously at a rate of 3% per annum. In Panel A, the BHLS value, the excess or difference between the risky and the risk-free values, goes from zero to 121.89. With conventional log or net returns, the BHLS path would have a total return of positive infinity, an analysis of which is out of the question. Excess return, the difference between the risky and the risk-free returns, can be computed coercively by ignoring the difference between the beginning-period capitals.[13] However, it is an open question whether the excess return series correctly describes the BHLS value process.

For illustration, let's start a path with 50 and compound on the excess log or net returns and end at 140.30 (Column 6, Panel A) or 142.21 (Column 8, Panel A), respectively. If we compound from 50 on the mean excess log or net return, 10.32% or 17.86%, we arrive at the final value of 140.30 or 258.67, respectively. Once again, the mean of log returns is path independent, while the mean of net returns significantly exaggerates the compounding effect.

Table 3 Here

One might be tempted to conclude that the excess log returns (Column 5, Panel A, Table 3) can be used to approximate the BHLS value series (Column 4, Panel A) because the final value of 140.30 is quite close to the final BHLS value of 121.89. Unfortunately, the results in Panel B invalidate this conclusion. The final BHLS value is negative (-18.66), but under log returns, the compounded final value is still positive (36.17). With one zero and five negative numbers in the BHLS series (Column 4, Panel B), returns can no longer be defined properly. As a result, we cannot use the excess log returns (Column 5, Panel B) to approximate the BHLS value series (Column 4, Panel B). Note that the mean excess log return is negative, so it gets the sign correct at least, while the positive excess net return has the wrong sign. Finally, if we subtract 50 from Column 6, Panel A, the final value is 90.30 for a second BHLS portfolio (i.e., long excess log return and short zero risk-free rate), which is not the same as the original BHLS final value of 121.89 (Column 4, Panel A). Obviously, the mean excess return underestimates the compounding effect.

[13] As pointed out previously, one cannot directly take the difference of two returns if the beginning-period capitals are unequal.

To formalize the above discussion, let's assume the mean log return of the market is higher than the (constant) risk-free rate. The final difference between the buy-and-hold market and the buy-and-hold risk-free portfolio is definitely larger than that of the buy-and-hold excess market and the buy-and-hold zero risk-free rate portfolio. It is easy to prove that:

$$e^{n\bar{y}} - e^{nr} > e^{n(\bar{y}-r)} - e^{0}, \qquad \bar{y} > r$$

because $e^{n\bar{y}} - e^{nr} = (e^{n(\bar{y}-r)} - e^{0})e^{nr}$ and $e^{nr} > 1$. A second proof is given in the Appendix. We have Proposition 2 as follows.

**Proposition 2** The difference between the buy-and-hold market and the buy-and-hold risk-free asset (MmTb) is larger than that between the hypothetical buy-and-hold excess market and the hypothetical buy-and-hold zero risk-free rate asset (XmFv).

Furthermore, $e^{n(\bar{y}+\delta)} - e^{n(r+\delta)} > e^{n\bar{y}} - e^{nr}, \delta > 0$. Therefore, Collonary 1 follows.

**Collonary 1** Given the same excess return, XmFv cannot explain the difference between a high interest rate regime with a higher MmTb and a low interest rate regime with a lower MmTb.

A few numbers can make the points more transparent. Say the typical market grows 10% per annum, and the risk-free rate is 3% per annum. In 20 years, 100 invested in the market (the Treasury bond) will grow to 739 (182), with an actual difference of 557 for MmTb, while the investment in the excess return of 7% (0%) will only grow to 406 (100), with a much smaller difference of 306 for XmFv. The actual difference (i.e., MmTb) is underestimated by 45% by XmFv, which is egregious.[14]

Again, an important point needs to be made about possible approximations with monthly returns that are widely used in asset pricing. Let's assume that there are 30 (365) days in a month (year), and we compound the daily returns, namely 356$^{th}$ of 10%, 3%, and 7%, for a month, XmFv turns out to underestimate the actual MmTb by 0.25%. A quarter of one percent may appear tiny. Still, unfortunately, we cannot ignore this tiny monthly difference, because over 20 years, which is relatively short in typical empirical asset pricing studies nowadays, the underestimation of MmTb by XmFv becomes enormous. Therefore, one should not ignore slight monthly differences indiscriminately.

To conclude, a return series cannot be well defined for the BHLS value series due to zero or negative values. Further, the BHLS market value process is not the same as that of the portfolio of long excess market and short zero risk-free rate, and the latter underestimates the compounding effects. The hypothetical excess market, which is commonly utilized in regressions, seems to be non-tradable, as does the hypothetical zero risk-free rate portfolio.

[14] With net returns, the investment in the excess return of 7.472% will grow to 423, because again the "average" of net returns, when compounded, overestimates the actual excess return. The actual difference is still underestimated by 42%.

Finally, let's summarize the results discussed in this section. First, the long-short (zero-cost) portfolio should be avoided by asset pricing studies because a meaningful definition of returns for its value process is impossible. Second, if regarded as portfolio returns approximately, log-return differences work better for both PRLS and BHLS. Third, the excess market return, which is commonly utilized in regressions, underestimates the compounding effect of the market compounded within a BHLS portfolio. These results lead to one drastic implication: the asset pricing factors would be better formulated on non-excess or non-difference log returns.

## 4. Asset Pricing Models

### 4.1. Single index model

Now, we analyze a single index or modified capital asset price model (CAPM). The return on a stock at time *t* can be expressed as follows:

$$\mu_t - r = \alpha + \beta(\mu_t^m - r) + \varepsilon_t \tag{5}$$

where $r$ is the (constant) risk-free interest rate, $\mu_t^m$ is the return on the market, and $\varepsilon_t$ is the random residual. Note that both $\mu_t$ and $\mu_t^m$ are computed from buy-and-hold price processes. As discussed in Section 3, the market factor, $\mu_t^m - r$, is either a PRLS or BHLS portfolio, with the borrowed cash or a risk-free bond as the short asset. Here a constant risk-free rate is used to make the exposition simpler.

From OLS regressions, the estimated coefficients are as follows (Elton and Gruber, 1987; Pindyck and Rubinfeld, 2000):

$$\hat{\beta} = \frac{\sum_t(\mu_t - \bar{\mu}_t)(\mu_t^m - \bar{\mu}_t^m)}{\sum_t(\mu_t^m - \bar{\mu}_t^m)^2} \tag{6}$$

$$\hat{\alpha} = \bar{\mu}_t - \hat{\beta}\bar{\mu}_t^m$$

where by the definition of OLS, $\bar{\mu}_t$ and $\bar{\mu}_t^m$ are arithmetic averages of returns on the stock (or portfolio) and the market, respectively.

Assuming that log returns are used in Equation (5), we would be dealing with geometric means of returns in Equation (6) as in Expression (1). With BHLS portfolios on both sides of Equation (5), the regression is thus well defined. As discussed in Section 3, however, excess returns, which lack rational economic meaning, underestimate the real terminal value if compounded. Therefore, we suggest that non-excess log returns should be employed in Equation (5).

If Equation (5) involves net returns, Equation (6) employs arithmetic means of returns as in Expression (2). With net returns, however, we are not dealing with the original buy-and-hold strategies. As discussed in Section 3, both the (excess) stock and the (excess) market might be regarded coercively as PRLS portfolios.

In addition to the issues discussed in Section 3, the regression with PRLS returns has three problems. First, periodic rebalancing in the long run is economically irrelevant because it faces the tiny size paradox (see Section 5.1 later).[15] Second, periodic rebalancing, by ignoring the compounding effect, will not be economically attractive for long-term investments. Third, the ignored extra cash flows in each period in periodic-rebalancing can dominate the investment returns (see Table 1 and Section 2.4).

Mathematically, (monthly) net returns can approximate (monthly) log returns. One may argue that the OLS of Model (5) with net returns can still approximate that with log returns. Unfortunately, the approximation can lead to egregious errors, as shown in Sections 2.4 and 3.4. Further, net returns exaggerate both alpha and its *t*-value of OLS (see Section 5.2 later) and even lead to pathological alphas and Sharpe ratios (see Section 5.3 later). To conclude, Model (5) is only correctly specified with buy-and-hold processes under non-excess log returns.

[15] The size of a periodic-rebalancing portfolio that starts with the US market capitalization in 1926 accounts only for 0.067% of the US market capitalization in 2020 (see Tables 4 and 5).

### 4.2. Fama-French three-factor

In addition to the market factor, the famous Fama-French three-factor (FF3) model utilizes the size (SMB) factor and the book-to-market or value (HML) factor (Fama and French, 1993). The time-series model is as follows:

$$\mu_t - r = \alpha + \beta(\mu_t^m - r) + \beta_s * SMB_t + \beta_v * HML_t + \varepsilon_t$$

In Fama and French (1993), the dependent variable is the excess return of one of 25 (dependent) portfolios formed from the intersections of five size and five book-to-market (BE/ME) quintiles.[16] Even though they might be interpreted as monthly rebalancing (i.e., periodic-rebalancing), the dependent portfolios are more meaningful when viewed as buy-and-hold in terms of real-world investing. Similarly, the market factor can be regarded as either PRLS or BHLS, as discussed in Section 3.

However, compared with the market factor, the SMB and HML factors are more complicated. In June of year *t*, Fama and French (1993) “construct six portfolios from the intersections of the two ME and the three BE/ME groups.” “Monthly value-weighted returns on the six portfolios are calculated from July of year *t* to June of *t*+1, and the portfolios are reformed in June of *t*+1.” Implicitly, net returns are used here because it is known that the net portfolio return over one period is the value-weighted average of the net returns of the portfolio's components.[17] Between yearly "reformulation," the six Fama-French portfolios can be either monthly rebalancing or buy-and-hold, however.

[16] ME (or size) is the market value of a firm, while BE is the book value of a firm.

Fama and French (1993) continue that "Our portfolio SMB (small minus big), meant to mimic the risk factor in returns related to size, is the difference, each month, between the simple average of the returns on the three small-stock portfolios (S/L, S/M, and S/H) and the simple average of the returns on the three big-stock portfolios (B/L, B/M, and B/H)" and "HML is the difference, each month, between the simple average of the returns on the two high-BE/ME portfolios (S/H and B/H) and the simple average of the returns on the two low-BE/ME portfolios (S/L and B/L)."[18]

Let's analyze HML because it is similar to SMB but involves only four portfolios. After a moment of reflection, it is clear that the S/H and B/H portfolios in HML have the same value weight or are of the same capital at the beginning of each month so that the net returns of the two portfolios at the end of each period can be averaged by simply dividing the sum of the two net returns by two. In addition, the long S/H-B/H composite and the short S/L-B/L composite in HML are also of equal capital, so HML is a zero-cost portfolio in each month, and HML over each period can be defined as the simple difference between the two averages of the composite returns. Therefore, the four portfolios (S/H, B/H, S/L, and B/L) utilized to compose HML are all of the equal amount of capital at the beginning of each month. Similarly, the six Fama-

[17] In this particular case, log return, which is nonlinear, cannot be utilized directly.
[18] HML: high minus low. All quotes are from Page 9 of Fama and French (1993).

French portfolios used to form SMB are all of the same amount of capital at the beginning of each month. Consequently, both SMB and HML are essentially PRLS portfolios.

As discussed in Section 3, the market factor is naturally BHLS, but may be coercively treated as PRLS. With the market as BHLS though, the FF3 model may suffer from misspecification. Because with a BHLS portfolio (market) that usually grows over time and two PRLS portfolios (size and value) that are fixed to the initial small size, the three factors are inherently inconsistent within the same regression. As pointed out earlier, the dependent variables are naturally buy-and-hold excess returns in nature (with compounding) and cannot be explained by PRLS factors that lack compounding. On the other hand, the FF3 model would be consistent with the market as PRLS. Unfortunately, periodic rebalancing ignores compounding and suffers severely from "the tiny size paradox," as pointed out previously.

## 5. Numerical Analyses

### 5.1. The tiny size paradox

As discussed earlier, the market factor in FF3 is more naturally buy-and-hold but can be periodic-rebalancing. If FF3 is formulated with PRLS portfolios, all three factors may suffer from the problem of the small "size of an effect" or economic irrelevancy (Nuzzo, 2014). K. French provides monthly returns online for the FF3 factors going back to July 1926. As an illustration, let's look at the US stock market at the end of June 1926 (Table 4).[19] The three partitions of the market in Table 4 are all quite skewed in terms of market value or capitalization; the “Small” size firms account for 5.7% of the market,[20] the “High” BE/ME firms are only 7.5% of the market, while the S/L firms sum up to a negligible percentage of 1.4%. By comparison, the three partitions of the market become even more skewed at the end of 2020 (Table 5), with Small at 3.3%, High at 7.9%, and S/H at 0.9% of the market, respectively. Note that the BE/ME breakpoints for 2020 are dramatically different from those of 1926, implying that the partition of the bottom 30%, middle 40%, and top 30% suggested in Fama and French (1993) seems to be somewhat arbitrary and is unstable in the long run. Similarly, the size breakpoint in 2020 is 176 times of that in 1926 and, more surprisingly, 1.7 times of the size of the biggest firm in 1926. It seems puzzling that the very “size” underlying the important size risk factor can itself vary so much, to say the least.

[19] Tables 4-6 are provided by Biao Yi.

Table 4 Here

In principle at least, the six portfolios used to compose the Fama-French factors, SMB and HML, should all be tradable.[21] Factor means are commonly reported (see, for example, Hou et al. (2015) and Liu et al. (2019)), which implies the tradability of factors. Remember that all six (four) portfolios in SMB (HML) have to be the same amount of capital, as discussed previously in Section 4.2. Because S/L, S/H, BL, and B/H appear in both SMB and HML, one-half of the minimum amount among these four portfolios would be one possible common capital for all the six Fama-French portfolios; that turns out to be $172 million for S/L in 1926 (Table 4). Even if we use all the available capital amount of S/L to trade, the capitalization of SMB (HML) only accounts for 4.2% (2.8%) of the US market. Economically, both SMB and HML are insignificant. Furthermore, the tradable portions of SMB and HML would be much smaller.[22] Finally, it is worth noting that the size factor within the *q*-factor model (Hou et al., 2015), defined as the difference between nine small-size and nine big-size portfolios, is even smaller and, with only a slight variation in formulation, becomes statistically distinct from SMB of FF3.[23] That the size factors in different models may lack statistical robustness may be a concern worthy of further study.

[20] For 1991, the small firms is about 8% of the market (Fama and French, 1993).

[21] Remember, Footnote 6 points out that the long-short zero-cost portfolio, namely SMB or HML, is not tradable in practice.

Table 5 Here

One might argue that the amount of capital does not matter because we use only returns in regression in empirical studies. But the amount of capital underlies the actual investments as well as the periodic-rebalancing long-short portfolio across time and, most importantly, determines the overall P&L of the investment. For example, the allocation of capital between SMB and HML mentioned in the previous paragraph can be arbitrary because the resultant factors, expressed in net-return differences, are unaffected by the actual amounts of capital. However, if 99 percent of capital were allocated to SMB, HML would not matter at all in real-world investments. This implies the hazardous nature of considering only returns in asset pricing studies.

The initial (small) size (i.e., the amount of capital) of the tradable portfolios,

---

[22] If one requires that the trading of the Fama-French factors does not move the market, the size of tradable portions of SMB and HML would be on the order of 100th of their available market capitalizations.
[23] Hou et al. (2015) report that the monthly mean return of their size factor is 0.31% (significant at the 5% level), while that of SMB is insignificant.

which has to be determined at the very beginning (of 1926), leads to a paradox for long-run time-series studies. With PRLS portfolios, if we start with the size of $172 million for S/L at the very first time point of a time series, we are stuck with it for all the subsequent time points, be it ten, twenty, or 100 years.[24] At the end of 2020, the US stock market has a size of $36 trillion with 3,107 firms (in CRSP with Share Code 10 or 11). With $172 million (the whole available capital for S/L in 1926) to begin with, the SMB and HML factors together only account for 0.0047% of the US market in 2020 and thus are economically irrelevant in 2020. We call the limitation on the initial amount of capital the tiny-size paradox.[25]

Worse, the initial size in 1926 would not have survived the Great Depression, even though it was pretty small already. By the middle of 1932, the S/L size dropped to 48 million (Table 6), less than 14% of the S/L size in 1926. This implies that a PRLS portfolio starting at any time may be impossible to maintain its initial amount of capital later in practice.

Table 6 Here

It might be argued that small firms tend to have higher percentage returns on average because they are only a small portion of the market; similarly, high BE/ME firms seem to show higher percentage returns on average, for they are also only a small portion of the market.[26] Practically speaking, such higher percentage returns cannot be scaled up by allocating more capital to small firms or high BE/ME firms. In other words, those higher percentage returns will not have a big effect on the overall capital gain of an investment when the size of the investment is large relative to the total size of small firms or high BE/ME firms. Lastly, the contrast between "Small" and "Big" may be misleading, and their match seems more extreme than David versus Goliath. In essence, we can only subtract a tiny portion of "Big" from "Small" so that SMB by definition does not have a big size as an effect, and so does HML.

---

[24] Arguably, an average of long run time-series returns fail to consider the time value of money, and therefore, may be questionable in itself.

[25] The equal-weighted portfolios, commonly utilized as robustness checks, is by definition even smaller. In 1926, the equal-weighted "market" portfolio has a tiny size of less than 18 million (i.e., 41 thousands, the size of the smallest firm, times 429, the number of firms), or 0.07% of the market (see Table 4). Further, it is unclear what the correspounding data generating process is, treated as either buy-and-hold or periodic-rebalancing.

[26] Fama and French (1993) write of "the negative relation between size and average return" and "the positive relation between BE/ME and average return" on Page 8.

In summary, focusing purely on net returns, as with the FF3 model, can be biased and misleading; what really matters in investing is in fact the magnitude of capital gains, not higher percentage returns on a tiny portion of the whole investment. In a Nature paper, Nuzzo (2014) argues that the right question to ask should be "[h]ow much of an effect is there." We believe that in asset pricing, we should also be asking the same question.

**5.2. Inflated alpha**

The inflation of alpha and its *t*-value by net returns can further be gauged by Monte Carlo simulations. Using a ten-year S&P 500 index as the market factor, we simulate buy-and-hold strategies under geometric Brownian motions (GBM). To be somewhat comprehensive, we take the S&P 500 index from January 1, 1990 to January 1, 2000 as the high-return market (with an annual arithmetic mean of returns of 15.4%), while from January 1, 2000 to January 1, 2010 as the low return market (with annual arithmetic mean of returns of -1.3%). Further, we simulate two drift rates, 8% and 16%. The volatilities are the same as those of the ten-year S&P 500 index for the chosen two decades.

Table 7 shows the numbers of significant alphas and betas. For the drift rate of 8%, 80 simulated buy-and-hold paths have a significant beta under log returns with the high return market as the factor (Panel A). As expected, there are more (413) paths with a significant beta if drift is higher at 16%. Further, *t*-values of alpha are inflated one level up by 41.3% and 21.8% for drift rates of 8% and 16%, respectively.

Table 7 here

These results are largely confirmed by the low return index as the factor (Panel B). The *t*-values of alphas are inflated one level up by 48.6% and 31.7% for drifts of 8% and 16%, respectively.

In summary, the utilization of net returns can overestimate the significance level of alphas by at least 20%, which is undoubtedly high and cannot be ignored lightly. These results imply that many documented anomalies or factors in the asset pricing literature might be artifacts of using net returns, which seems to corroborate Harvey et al. (2016) from a different perspective.

**5.3. Pathological performance measures**

Net returns may lead to erroneous performance measures for buy-and-hold strategies, as first reported by Guo and Liu (2019). Following exactly their approach, we obtain two time series (Table 8 and Figure 1). Both price series start at 50, but the first (Up) series goes up to 133.94 after 120 months (or ten years), while the second (Down) series is down to 26.10. For the whole sample period, the Down series is above the Up series only in the third and fifth months. The Down series drops to 2.11 in the 34$^{th}$ month from 27.91 in the 33rd month and remains below ten until the 94th month. Clearly, the Up buy-and-hold time series outperforms the Down series.

Table 8 here

Figure 1 here

Net returns yield egregious errors for the average returns and performance measures in Table 9, however, as similarly shown in Guo and Liu (2019). The Down series, with a big loss over the sample period, shows a huge monthly average net return of 6.1%, which is outrageously wrong and higher than the average net return of 1.1% for the Up series. This erroneous average net return leads directly to the wrong performance measures. The Down series has a Sharpe ratio of 0.188, which is higher than that of the Up series, contradicting the simple fact of a total loss for Down but a total gain for Up. Finally, the Down series shows a positive alpha at the 10% significance level against the Up, which is once again wrong.

Table 9 here

The average log returns for the two series are of the correct signs, positive and negative, respectively. Further, Sharpe ratios are of the correct signs, positive and negative, respectively. Finally, the alpha of the Down series has the correct negative sign, even though it is not significant. In summary, average returns, Sharpe ratios, and alphas computed with log returns are consistent with the intuition gained directly from Tables 8 and 9 and Figure 1.

To conclude, the Sharpe ratio should be based on log returns, which seems fortuitous to address Scholes' concern with the conventional Sharpe ratio that is based on net returns (see the Introduction). Further, Proposition 2 states that the excess returns under-estimate the compounding effects, compared to holding the market and the risk-free assets separately. Therefore, we suggest that the Sharpe ratio should be defined by non-excess log returns instead of excess net returns.

## 6. Conclusions

In this paper, we analyze the value or data-generating processes of buy-and-hold, periodic-rebalancing, and zero-cost portfolios. Buy-and-hold can be modeled by the geometric Brownian motion, but periodic rebalancing cannot. We show that the arithmetic average of log returns, which we call the geometric mean of returns and accounts for the compounding effect, correctly measures the time-series average behavior of the buy-and-hold strategy. In contrast, the arithmetic average of net returns, which is termed the arithmetic mean of returns and lacks compounding, does not describe even the periodic-rebalancing strategy and is less useful. Numerically, we show that the ignored cash flows in each period if compounded at a risk-free rate, dominate the simple sum of returns of the periodic-rebalancing investing.

Our detailed analyses of long-short (zero-cost) portfolios show that its value process does not allow a meaningful definition for returns. Further, the market excess return or simply the market factor, which is widely utilized in regressions, significantly underestimates the compounding effect of the market within a buy-and-hold long-short (BHLS) portfolio. We argue that the long-short (zero-cost) portfolio should be avoided by asset pricing studies and suggest that the asset pricing factors would be better represented by non-excess or non-difference log returns.

Because the OLS regression of time series is in essence based on arithmetic averages, we argue that log returns should be used for single index models, which are buy-and-hold portfolios by definition. On the other hand, using net returns leads to either inflated alphas or uninteresting periodic-rebalancing strategies. For the same reason, the market factor and Sharpe ratio are better defined with non-excess log returns.

For the Fama-French three-factor (FF3) model, the size and value factors are periodic-rebalancing long-short (PRLS) portfolios. The market factor can arguably be either BHLS or PRLS. If the market were BHLS, the three factors would be inconsistent, and FF3 may be misspecified. With the market as PRLS, FF3, suffering from the tiny size paradox and lacking compounding, may be economically unattractive and irrelevant.

Overall, the paper has wide implications for studies of time-series empirical asset pricing. Our analyses of FF3 apply to other FF3-type models, such as the Fama-French five-factor (Fama and French, 2015), the $q$-factor model (Hou et al., 2015), and the Chinese three- and four-factors (Liu et al., 2019), among others. Our arguments imply that many documented anomalies or factors in the asset pricing literature might be artifacts of using differences in net returns.

## Appendix

### A Second Proof of Proposition 2

Assume that $y > x \geq 0$. By Taylor expansion, we have:

$$e^y - e^x = (y-x) + \frac{1}{2}(y^2 - x^2) + \frac{1}{6}(y^3 - x^3) + \cdots$$

$$e^{(y-x)} - e^0 = (y-x) + \frac{1}{2}(y-x)^2 + \frac{1}{6}(y-x)^3 + \cdots$$

First, comparing the quadratic terms, we obtain:

$$(y-x)^2 < (y-x)(y+x) = y^2 - x^2$$

For the cubic terms, we multiply the quadratic result by $(y-x)$:

$$(y-x)^3 < (y-x)(y^2 - x^2)$$

$$= (y^3 - x^3) + x(x^2 - y^2) + x^2(x-y)$$

Clearly, the second and third terms in the last expression are both negative, and thus, we have:

$$(y-x)^3 < y^3 - x^3$$

Finally, let's assume:

$$(y-x)^{n-1} < y^{n-1} - x^{n-1}$$

and multiply both sides by $(y-x)$:

$$(y-x)^n < (y-x)(y^{n-1} - x^{n-1})$$

$$= (y^n - x^n) + x(x^{n-1} - y^{n-1}) + x^{(n-1)}(x-y)$$

Again, the second and third terms in the last expression are both negative. Therefore:

$$(y-x)^n < y^n - x^n$$

By induction, we have established the inequality for all $n$. Therefore, we conclude that:

$$e^y - e^x > e^{(y-x)} - e^0$$

## References

Cochrane, John H., 2005, Asset pricing, rev. ed. (Princeton University Press, Princeton).

Elton, Edwin J., and Martin J. Gruber, 1987, Modern portfolio theory and investment analysis, 3rd ed. (John Wiley & Sons, New York).

Fama, Eugene F., and Kenneth R. French, 1993, Common risk factors in the returns on stocks and bonds, *Journal of Financial Economics* 33, 3-56.

Fama, Eugene F., and Kenneth R. French, 2015, A five-factor asset pricing model, *Journal of Financial Economics* 116, 1-22.

Guo, Shuxin, and Qiang Liu, 2019, Volatility-managed portfolios: True market-timing with a false theory? *SSRN*, dx.doi.org/10.2139/ssrn.3385377.

Guo, Shuxin, and Qiang Liu, 2022, Is the annualized compounded return of Medallion over 35%? *SSRN*, dx.doi.org/10.2139/ssrn.4174685.

Harvey, Campbell R., Yan Liu, and Heqing Zhu, 2016, ... and the cross-section of expected returns, *Review of Financial Studies* 29, 5-68.

Hou, Kewei, Chen Xue, and Lu Zhang, 2015, Digesting anomalies: An investment approach, *Review of Financial Studies* 28, 650-705.

Hull, John C., 2015, *Options, Futures, and Other Derivatives*, 9th ed. (Prentice Hall, Upper Saddle River, NJ).

Liu, Jianan, Robert F. Stambaugh, and Yu Yuan, 2019, Size and value in China, *Journal of Financial Economics* 134, 48-69.

Merton, Robert C., 1992, Continuous-time finance, rev. ed. (Blackwell Publishers, Malden, Massachusetts).

Moreira, Alan, and Tyler Muir, 2017, Volatility-managed portfolios, *Journal of Finance* 72, 1611-1644.

Nuzzo, Regina, 2014, Scientific method: Statistical errors, *Nature* 506, 150-152.

Pindyck, Robert, and Daniel L. Rubinfeld, 2000, Econometric models and economic forecasts, 4th ed. (McGraw-Hill Education, New York).

Sharpe, William F., 1964, Capital asset prices: A theory of market equilibrium under conditions of risk, *Journal of Finance* 19, 425-442.

**Table 1. Monthly S&P 500 index from Jan. 1, 1960 to Jan. 1, 2020**

$\bar{y}$: mean of monthly log returns. $\bar{\mu}$: mean of monthly net returns. $\mu_t$: month-*t* net return.
Risk-free rate (rf) is 3% per annum. 1-month T-Bill: from the data library of K. French.

| | |
|---|---|
| Number of months ($N$) | 720 |
| $S_0$ (Jan. 1, 1960) | 55.61 |
| $S_T$ (Jan. 1, 2020) | 3225.52 |
| *Log Returns* | |
| Geometric-mean ($\bar{y}$) | 5.640E-03 |
| Standard deviation | 4.224E-02 |
| Sharpe ratio | 7.440E-02 |
| $S_0 e^{\bar{y}N}$ | 3225.52 |
| *Net Returns* | |
| Arithmetic-mean ($\bar{\mu}$) | 6.543E-03 |
| Standard deviation | 4.196E-02 |
| Sharpe ratio | 9.636E-02 |
| $S_0(1+\bar{\mu})^N$ | 6089.20 |
| *Net Returns: Hypothetical Periodic-Rebalancing* | |
| Sum of monthly P&L: $S_0\bar{\mu}N$ | 261.99 |
| Final value: $S_0(1+\bar{\mu}N)$ | 317.60 |
| Sum: $S_0\mu_t$ compounded to *T* (with rf) | 674.17 |
| Sum: $S_0\mu_t$ compounded to *T* (with 1-month T-Bill) | 951.20 |
| Sum: $S_0\bar{\mu}$ compounded to *T* (with rf) | 733.01 |
| Sum: $S_0\bar{\mu}$ compounded to *T* (with 1-month T-Bill) | 1170.24 |

**Table 2. Hypothetical long, short, and buy-and-hold long-short (BHLS) portfolios over two periods**

For BHLS, the Value and Net/Log returns are the differences between those of the Long and Short legs, respectively. Mean returns are the simple average of net/log returns over the two periods.

| *A. No-Gain for BHLS* | | | | | Mean (%) |
|---|---|---|---|---|---|
| Long | Value | 100 | 150 | 100 | |
| | Net return (%) | | 50.0 | -33.3 | 8.3 |
| | Log return (%) | | 40.5 | -40.5 | 0.0 |
| Short | Value | 100 | 80 | 100 | |
| | Net return (%) | | -20.0 | 25.0 | 2.5 |
| | Log return (%) | | -22.3 | 22.3 | 0.0 |
| BHLS | Value | 0 | 70 | 0 | |
| | Net return (%) | | 70.0 | -58.3 | 5.8 |
| | Log return (%) | | 62.9 | -62.9 | 0.0 |
| *B. Gain for BHLS* | | | | | |
| Long | Value | 100 | 90 | 115 | |
| | Net return (%) | | -10.0 | 27.8 | 8.9 |
| | Log return (%) | | -10.5 | 24.5 | 7.0 |
| Short | Value | 100 | 10 | 70 | |
| | Net return (%) | | -90.0 | 600.0 | 255.0 |
| | Log return (%) | | -230.3 | 194.6 | -17.8 |
| BHLS | Value | 0 | 80 | 45 | |
| | Net return (%) | | 80.0 | -572.2 | -246.1 |
| | Log return (%) | | 219.7 | -170.1 | 24.8 |
| *C. Loss for BHLS* | | | | | |
| Long | Value | 100 | 10 | 70 | |
| | Net return (%) | | -90.0 | 600.0 | 255.0 |
| | Log return (%) | | -230.3 | 194.6 | -17.8 |
| Short | Value | 100 | 40 | 80 | |
| | Net return (%) | | -60.0 | 100.0 | 20.0 |
| | Log return (%) | | -91.6 | 69.3 | -11.2 |
| BHLS | Value | 0 | -30 | -10 | |
| | Net return (%) | | -30.0 | 500.0 | 235.0 |
| | Log return (%) | | -138.6 | 125.3 | -6.7 |

**Table 3. Simulated buy-and-hold long-short (BHLS) portfolios**

BHLS: the difference between the risky and risk-free values. Excess: the difference between the risky and risk-free log/net returns. Comp: the value compounded on the Excess. Mean: the simple average of net/log returns over ten years.

| | | | | Log Return | | Net Return | |
|---|---|---|---|---|---|---|---|
| Year | Risky | Risk-free | BHLS | Excess | Comp | Excess | Comp |
| *A.* | | | | | | | |
| 0 | 50.00 | 50.00 | 0.00 | | 50.00 | | 50.00 |
| 1 | 86.98 | 51.52 | 35.46 | 0.5236 | 84.41 | 0.7091 | 85.46 |
| 2 | 174.36 | 53.09 | 121.27 | 0.6655 | 164.20 | 0.9742 | 168.70 |
| 3 | 228.17 | 54.71 | 173.46 | 0.2390 | 208.53 | 0.2782 | 215.63 |
| 4 | 207.84 | 56.37 | 151.47 | -0.1233 | 184.34 | -0.1195 | 189.86 |
| 5 | 156.57 | 58.09 | 98.48 | -0.3133 | 134.76 | -0.2772 | 137.24 |
| 6 | 189.93 | 59.86 | 130.07 | 0.1632 | 158.65 | 0.1827 | 162.30 |
| 7 | 206.48 | 61.68 | 144.80 | 0.0535 | 167.37 | 0.0567 | 171.50 |
| 8 | 303.61 | 63.56 | 240.05 | 0.3555 | 238.83 | 0.4400 | 246.95 |
| 9 | 284.98 | 65.50 | 219.49 | -0.0933 | 217.55 | -0.0918 | 224.28 |
| 10 | 189.38 | 67.49 | 121.89 | -0.4387 | 140.30 | -0.3659 | 142.21 |
| Mean | | | | 0.1032 | | 0.1786 | |
| Final value compounded on the mean | | | | | 140.30 | | 258.67 |
| *B.* | | | | | | | |
| 0 | 50.00 | 50.00 | 0.00 | | 50.00 | | 50.00 |
| 1 | 53.41 | 51.52 | 1.89 | 0.0361 | 51.84 | 0.0378 | 51.89 |
| 2 | 51.36 | 53.09 | -1.73 | -0.0692 | 48.37 | -0.0689 | 48.32 |
| 3 | 100.21 | 54.71 | 45.50 | 0.6384 | 91.58 | 0.9206 | 92.80 |
| 4 | 72.30 | 56.37 | 15.93 | -0.3564 | 64.13 | -0.3089 | 64.13 |
| 5 | 55.17 | 58.09 | -2.92 | -0.3004 | 47.49 | -0.2674 | 46.98 |
| 6 | 56.56 | 59.86 | -3.30 | -0.0051 | 47.25 | -0.0053 | 46.74 |
| 7 | 66.28 | 61.68 | 4.60 | 0.1286 | 53.73 | 0.1414 | 53.34 |
| 8 | 67.31 | 63.56 | 3.75 | -0.0146 | 52.95 | -0.0149 | 52.55 |
| 9 | 54.24 | 65.50 | -11.26 | -0.2460 | 41.40 | -0.2247 | 40.74 |
| 10 | 48.83 | 67.49 | -18.66 | -0.1350 | 36.17 | -0.1301 | 35.44 |
| Mean | | | | -0.0324 | | 0.0080 | |
| Final value compounded on the mean | | | | | 36.17 | | 54.12 |

## Table 4. Descriptions of the U.S. stock market and the six Fama-French portfolios at the end of June 1926

Min: minimum of a portfolio. Max: maximum of a portfolio. "Size" stands for the market equity or market capitalization. Market: all firms in CRSP with share code 10 or 11. Small/Big (size): NYSE median size of $14.3 million as the breakpoint. Low/Medium/High (book-to-market): NYSE BE/ME breakpoints of 0.706 and 1.550. S/L: Small and Low. S/M: Small and Medium. S/H: Small and High. B/L: Big and Low. B/M: Big and Medium. B/H: Big and High.

| | Number of Firms | Min Size (in $1,000) | Max Size (in $1,000) | Portfolio Size (in $1,000) |
|---|---|---|---|---|
| Market | 429 | 41 | 1,510,460 | 24,337,821 |
| Small | 215 | 41 | 14,281 | 1,381,116 |
| Big | 214 | 14,300 | 1,510,460 | 22,956,705 |
| Low | 129 | 810 | 822,168 | 9,822,388 |
| Medium | 172 | 1,088 | 1,510,460 | 12,692,805 |
| High | 128 | 41 | 184,186 | 1,822,628 |
| S/L | 44 | 810 | 13,275 | 344,681 |
| S/M | 72 | 1,088 | 14,100 | 520,663 |
| S/H | 99 | 41 | 14,281 | 515,772 |
| B/L | 85 | 14,375 | 822,168 | 9,477,707 |
| B/M | 100 | 14,616 | 1,510,460 | 12,172,141 |
| B/H | 29 | 14,300 | 184,186 | 1,306,856 |

**Table 5. Descriptions of the U.S. stock market and the six Fama-French portfolios at the end of December 2020**

Min: minimum of a portfolio. Max: maximum of a portfolio. “Size” stands for the market equity or market capitalization in US dollars. Market: all firms in CRSP with share code 10 or 11. Small/Big (size): NYSE median size of $2,515 million as the breakpoint. Low/Medium/High (book-to-market): NYSE BE/ME breakpoints of 0.309 and 0.770. S/L: Small and Low. S/M: Small and Medium. S/H: Small and High. B/L: Big and Low. B/M: Big and Medium. B/H: Big and High. Size is grouped by the end of June 2020 market values. Book-to-market is partitioned by BE/ME computed from the end of 2019 fiscal year book values and the end of 2019 market values.

| | Number of Firms | Min Size ($million) | Max Size ($million) | Portfolio Size ($million) |
|---|---|---|---|---|
| Market | 3,107 | 7 | 2,255,969 | 36,289,470 |
| Small | 2,010 | 7 | 2,507 | 1,182,283 |
| Big | 1,097 | 2,516 | 2,255,969 | 35,107,187 |
| Low | 1,049 | 7 | 2,255,969 | 24,584,635 |
| Medium | 1,139 | 10 | 387,335 | 8,842,295 |
| High | 919 | 7 | 543,615 | 2,862,540 |
| S/L | 493 | 7 | 2,507 | 342,301 |
| S/M | 750 | 10 | 2,506 | 512,598 |
| S/H | 767 | 7 | 2,492 | 327,384 |
| B/L | 556 | 2,516 | 2,255,969 | 24,242,334 |
| B/M | 389 | 2,523 | 387,335 | 8,329,697 |
| B/H | 152 | 2,536 | 543,615 | 2,535,156 |

**Table 6. Descriptions of the U.S. stock market and the six Fama-French portfolios at the end of June 1932**

Min: minimum of a portfolio. Max: maximum of a portfolio. "Size" stands for the market equity or market capitalization in US dollars. Market: all firms in CRSP with share code 10 or 11. Small/Big (size): NYSE medium size of $2.4 million as the breakpoint. Low/Medium/High (book-to-market): NYSE BE/ME breakpoints of 1.843 and 6.320. S/L: Small and Low. S/M: Small and Medium. S/H: Small and High. B/L: Big and Low. B/M: Big and Medium. B/H: Big and High.

| | Number of Firms | Min Size (in $1,000) | Max Size (in $1,000) | Portfolio Size (in $1,000) |
|---|---|---|---|---|
| Market | 594 | 31 | 1,434,334 | 10,973,989 |
| Small | 297 | 31 | 2,419 | 280,420 |
| Big | 297 | 2,452 | 1,434,334 | 10,693,570 |
| Low | 179 | 75 | 1,434,334 | 8,350,717 |
| Medium | 237 | 121 | 218,057 | 2,214,281 |
| High | 178 | 31 | 69,915 | 408,991 |
| S/L | 30 | 75 | 2,419 | 47,723 |
| S/M | 120 | 121 | 2,419 | 135,152 |
| S/H | 147 | 31 | 2,371 | 97,545 |
| B/L | 149 | 2,576 | 1,434,334 | 8,302,994 |
| B/M | 117 | 2,452 | 218,057 | 2,079,129 |
| B/H | 31 | 2,472 | 69,915 | 311,446 |

**Table 7. Simulating single index model under geometric Brown motion**

Numbers in the table are instances for various cases defined by *t*-values, if not specified otherwise. Annual mean: annualized average of monthly net returns on S&P 500 index. $t_L(beta)$: *t*-value for the beta estimated from log returns. $t_L$: *t*-value for the alpha estimated from log returns. $t_N$: *t*-value for the alpha estimated from net returns. The risk-free rate is 3% per annum. Simulation details: the volatility is the same as that of S&P 500 index; the annual drift rate is either 8% or 16%; 10,000 price paths are simulated.

| | Annual drift rate | |
|---|---|---|
| | 8% | 16% |
| *A. Monthly S&P 500 (Jan. 1, 1990 to Jan. 1, 2000)* | | |
| Annual mean 15.4%; Annual volatility: 13.3% | | |
| $t_L(beta) > 1.66$ | 80 | 413 |
| $t_L < 1.66$ & $t_N > 1.66$ | 17 | 22 |
| $t_L < 1.98$ & $t_N > 1.98$ | 11 | 25 |
| $t_L < 2.62$ & $t_N > 2.62$ | 5 | 29 |
| $t_L < 3.37$ & $t_N > 3.37$ | 0 | 14 |
| Percentage (%) | 41.3 | 21.8 |
| *B. Monthly S&P 500 (Jan. 1, 2000 to Jan. 1, 2010)* | | |
| Annual mean -1.3%; Annual volatility: 16.4% | | |
| $t_L(beta) > 1.66$ | 173 | 423 |
| $t_L < 1.66$ & $t_N > 1.66$ | 38 | 32 |
| $t_L < 1.98$ & $t_N > 1.98$ | 27 | 36 |
| $t_L < 2.62$ & $t_N > 2.62$ | 16 | 40 |
| $t_L < 3.37$ & $t_N > 3.37$ | 3 | 26 |
| Percentage (%) | 48.6 | 31.7 |

**Table 8**. **Two monthly price series**

The initial price is 50 for both series. Up: first price series. Down: second price series.

| Month | Up | Down | Month | Up | Down | Month | Up | Down |
|---|---|---|---|---|---|---|---|---|
| 1 | 53.78 | 53.78 | 41 | 56.21 | 2.18 | 81 | 67.15 | 0.59 |
| 2 | 56.17 | 55.41 | 42 | 60.65 | 2.35 | 82 | 67.79 | 0.63 |
| 3 | 56.60 | 58.37 | 43 | 62.18 | 2.43 | 83 | 69.37 | 1.20 |
| 4 | 56.61 | 53.55 | 44 | 62.34 | 2.43 | 84 | 70.27 | 1.64 |
| 5 | 57.54 | 60.02 | 45 | 65.81 | 2.81 | 85 | 70.42 | 1.59 |
| 6 | 57.08 | 45.86 | 46 | 68.64 | 2.98 | 86 | 71.88 | 1.82 |
| 7 | 56.26 | 29.87 | 47 | 70.13 | 3.60 | 87 | 71.89 | 1.74 |
| 8 | 53.76 | 28.72 | 48 | 69.16 | 3.10 | 88 | 73.82 | 2.61 |
| 9 | 51.14 | 27.57 | 49 | 63.50 | 2.15 | 89 | 75.15 | 3.95 |
| 10 | 52.75 | 28.16 | 50 | 57.83 | 2.06 | 90 | 77.00 | 4.69 |
| 11 | 57.47 | 30.66 | 51 | 58.96 | 2.08 | 91 | 80.38 | 5.83 |
| 12 | 52.13 | 27.26 | 52 | 59.00 | 2.08 | 92 | 83.05 | 6.74 |
| 13 | 54.59 | 28.09 | 53 | 52.38 | 1.70 | 93 | 82.67 | 4.14 |
| 14 | 56.07 | 28.54 | 54 | 48.82 | 1.65 | 94 | 84.80 | 7.31 |
| 15 | 52.59 | 21.53 | 55 | 37.62 | 1.33 | 95 | 87.66 | 13.86 |
| 16 | 58.18 | 24.91 | 56 | 40.60 | 1.37 | 96 | 88.46 | 14.99 |
| 17 | 57.09 | 24.78 | 57 | 46.30 | 1.40 | 97 | 89.14 | 16.05 |
| 18 | 55.49 | 24.52 | 58 | 44.95 | 1.38 | 98 | 91.13 | 17.99 |
| 19 | 54.68 | 24.32 | 59 | 48.46 | 1.45 | 99 | 93.74 | 22.57 |
| 20 | 44.90 | 22.29 | 60 | 48.85 | 1.46 | 100 | 88.17 | 3.91 |
| 21 | 32.01 | 20.68 | 61 | 49.25 | 1.56 | 101 | 86.43 | 3.62 |
| 22 | 28.42 | 20.28 | 62 | 49.42 | 1.61 | 102 | 89.97 | 3.68 |
| 23 | 30.25 | 20.45 | 63 | 49.44 | 1.61 | 103 | 90.82 | 3.72 |
| 24 | 39.73 | 21.73 | 64 | 50.59 | 1.78 | 104 | 93.19 | 4.03 |
| 25 | 43.64 | 22.85 | 65 | 51.71 | 2.13 | 105 | 97.77 | 4.25 |
| 26 | 44.53 | 23.26 | 66 | 52.66 | 3.75 | 106 | 100.44 | 4.52 |
| 27 | 46.44 | 24.40 | 67 | 52.48 | 3.51 | 107 | 104.32 | 10.52 |
| 28 | 47.77 | 25.99 | 68 | 48.84 | 2.60 | 108 | 108.56 | 12.75 |
| 29 | 47.13 | 12.75 | 69 | 47.88 | 2.50 | 109 | 112.48 | 16.94 |
| 30 | 49.55 | 17.73 | 70 | 50.71 | 2.59 | 110 | 114.05 | 19.49 |
| 31 | 51.46 | 19.51 | 71 | 51.71 | 2.65 | 111 | 113.26 | 18.14 |
| 32 | 52.94 | 21.75 | 72 | 50.33 | 1.63 | 112 | 106.51 | 13.46 |
| 33 | 54.71 | 27.91 | 73 | 55.22 | 1.97 | 113 | 110.70 | 14.26 |
| 34 | 51.98 | 2.11 | 74 | 56.57 | 2.02 | 114 | 120.26 | 21.70 |
| 35 | 55.02 | 2.33 | 75 | 58.41 | 2.34 | 115 | 117.95 | 20.26 |
| 36 | 55.05 | 2.32 | 76 | 60.17 | 2.70 | 116 | 125.51 | 23.05 |
| 37 | 51.13 | 2.09 | 77 | 61.72 | 2.95 | 117 | 119.46 | 19.51 |
| 38 | 45.36 | 1.81 | 78 | 61.78 | 2.80 | 118 | 129.39 | 23.26 |
| 39 | 54.47 | 2.10 | 79 | 63.21 | 5.86 | 119 | 132.45 | 24.51 |
| 40 | 54.75 | 2.11 | 80 | 62.71 | 0.37 | 120 | 133.94 | 26.10 |

**Table 9. Statistics and performance measures for two price series**

Up series: Up or first series in Table 8. Down series: Down or second series in Table 8. SD: standard deviation. Sharpe: Sharpe ratio. Alpha is the intercept from regressing the Down series on the Up series. For computing the Sharpe ratio, the monthly risk-free rate of 0.25% is used for both net and log returns.

| | Mean | SD | Sharpe | Alpha *t*-value |
|---|---|---|---|---|
| *A. Net Returns* | | | | |
| Up series | 0.011 | 0.069 | 0.119 | |
| Down series | 0.061 | 0.308 | 0.188 | 1.70* |
| *B. Log Returns* | | | | |
| Up series | 0.008 | 0.070 | 0.081 | |
| Down series | -0.005 | 0.446 | -0.018 | -0.46 |

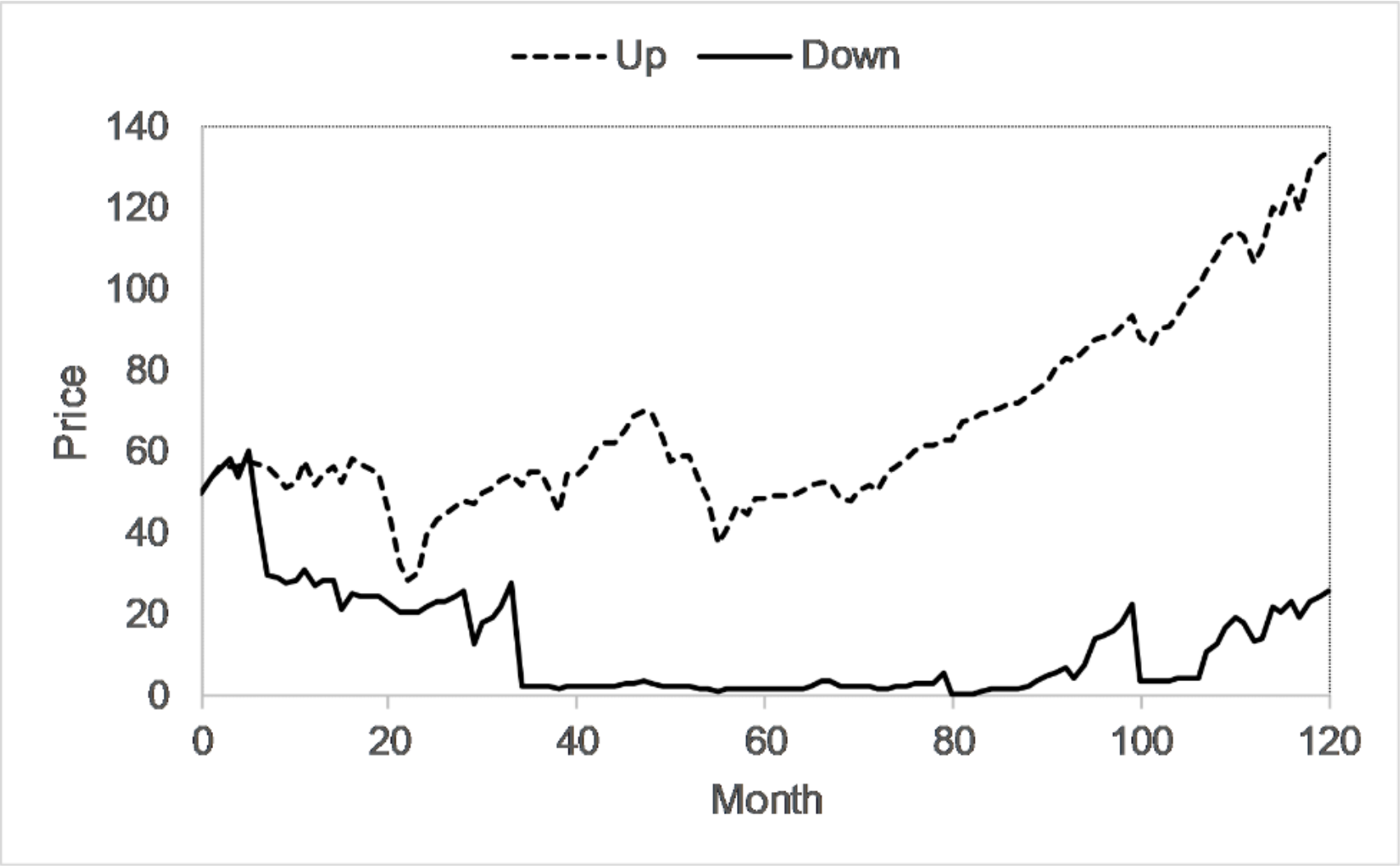

**Figure 1. Two monthly price series from Table 8**